\documentclass[prc,twocolumn,nofootinbib]{revtex4-2}

\usepackage{color}
\usepackage{graphicx}
\usepackage{dcolumn}
\newcolumntype{d}[1]{D{.}{.}{#1}}
\usepackage{bm}
\usepackage{braket}
\usepackage{amsmath}
\usepackage{here}
\usepackage{color}
\usepackage{multirow}

\usepackage{array}
\usepackage{enumerate}
\usepackage{amssymb}
\usepackage{bbm}
\usepackage{dsfont}
\usepackage[scr=rsfs]{mathalpha}
\usepackage{longtable}

\def\be{\begin{equation}} 
\def\ee{\end{equation}}

\newcommand{\vecbold}[1]{\mbox{\boldmath $#1$}}
\usepackage{xcolor}
\graphicspath{{figs/}}
\usepackage[bookmarks=true,colorlinks,
            citecolor=blue,linkcolor=blue,anchorcolor=blue,filecolor=blue,urlcolor=blue,
           ]{hyperref}

\usepackage[normalem]{ulem}

\renewcommand\sout{\bgroup\markoverwith
{\textcolor[rgb]{1,0.75,0.8}{\rule[.5ex]{2pt}{0.8pt}}}\ULon}
\begin{document}

\title{Application of the shift-invert Lanczos algorithm to 
a non-equilibrium Green function for transport problems}

\author{K. Uzawa}
\affiliation{ 
Department of Physics, Kyoto University, Kyoto 606-8502,  Japan} 

\author{K. Hagino}
\affiliation{ 
Department of Physics, Kyoto University, Kyoto 606-8502,  Japan} 

\begin{abstract}
Non-equilibrium Green's function theory and related methods are widely
used to describe transport phenomena in many-body systems, but they often require
a costly inversion of a large matrix.  We show here that the 
shift-invert Lanczos method can dramatically reduce the computational
effort.  We apply the method to two test problems, namely a simple
model Hamiltonian and to a more realistic Hamiltonian for nuclear fission.
For a Hamiltonian of dimension 66103 we find that the computation
time is reduced by a factor of 33 compared to the direct calculation
of the Green's function.
\end{abstract}

\maketitle

\section{introduction}

To describe transport phenomena in fermionic many-body systems is one of the most important topics 
in many fields of physics and chemistry \cite{datta_1995,Datta_2005}. 
For this purpose, the non-equilibrium Green function (NEGF) and related methods   \cite{DANIELEWICZ1984239, balzer2012nonequilibrium, Nonequilibrium_text} have been widely employed 
as an efficient computational method. 
For instance, 
they have been applied to calculate electronic properties of nanodevices, mesoscopic systems,  
and molecular junctions \cite{Nardelli1999, Taylor2001, Damle2001, Brandbyge2002, Xue2002, Thoss2018}. 
Recently, the NEGF method has been utilized also to describe nuclear fission reactions as well 
\cite{Alhassid2020, Bertsch2022, Weidenmuller2022, Bertsch2023, Uzawa2023}.

In this method, 
one assumes that a system can be divided into 
two external and an internal systems, 
and then constructs an effective Hamiltonian for the internal part, 
in which the couplings to the external parts are taken into account 
as self-energies. 
In the cases of electronic devices, the external parts correspond to the semi-infinite 
electrodes while the internal part is a nano device itself.
On the other hand, in the case of nuclear fission reaction, 
the external parts correspond to decay channels  
of a nucleus, 
while the internal part is for 
mono-nucleus states 
during a shape evolution of a nucleus. 
Due to the couplings to the continuum external states, 
the resultant effective Hamiltonian for the internal part 
becomes non-Hermitian, 
and many-body current is allowed to flow between the input and output channels.
The transition probability between the channels 
is calculated using the Green function with the effective Hamiltonian. 

While this method is applicable even for cases beyond the linear response theory,
the NEGF method requires a significant computational cost, especially for calculations of a Green function, which 
can be constructed by inverting a Hamiltonian matrix. 
Notice that 
the dimension of the Hamiltonian matrix considerably increases with the system size. 
In the case of 
silicon nanowire transistors in 
three-dimension,  
the matrix dimension becomes orders of $10^5$ \cite{Luisier2009}. 
As another example, 
we mention that the dimension is estimated to be orders of $10^6$ 
for neutron induced fission of actinide nuclei  
\cite{Bertsch2023}.

To reduce the numerical costs, specific matrix structures can be useful.
For example, under the tight-binding approximation, the Hamiltonian matrix becomes block-tridiagonal, and  
its inversion can be evaluated by inverting the sub-matrices, instead of the whole matrix  
\cite{Dy1979, Godfrin1991, Svizhenko2002, Petersen2008}. This method was applied e.g., in Ref. \cite{Luisier2009} by dropping off 
long-range off-diagonal elements due to the electron-phonon coupling. 
Notice that the applicability of this method is limited by the structure of the considered 
Hamiltonian. For instance, in the problem discussed in Ref. \cite{Luisier2009}, the Hamiltonian is no longer 
block-tridiagonal when one takes into account  
the long-range off-diagonal elements due to the electron-phonon coupling.  
In the case of nuclear induced fission, a Hamiltonian matrix is in general not block-tridiagonal \cite{Bertsch2023}. 

To overcome the drawback of the NEGF method,  
we here discuss the applicability of the Lanczos algorithm \cite{Lanczos1, Lanczos2, Lanczos_text}.
The standard Lanczos algorithm provides an efficient method to obtain the ground state 
of a large matrix. 
On the other hand, 
in the NEGF approach, the Green function at various energies has to be evaluated: for which 
one needs information 
on the eigenstates in the middle of the spectrum of a Hamiltonian matrix, rather than the ground state.  
Therefore the usual Lanczos algorithm is not efficient, and we will instead 
introduce the shift-invert Lanczos 
algorithm, in which an original eigenvalue equation is transformed to calculate efficiently  
states in the middle of the spectrum. 
In this way, the calculation of the NEGF method is accelerated by 
obtaining specific eigensolutions of a Hamiltonian matrix.

This paper is organized as follows. 
In Sec. \ref{formulation}, we summarize 
the non-equilibrium Green function method and  
introduce the shift-invert Lanczos method in the context of the NEGF method. 
In Sec. \ref{results}, we numerically demonstrate the efficiency of our approach for a simple schematic model 
as well as for a more realistic Hamiltonian for fission of $^{236}$U based on 
the density functional theory(DFT).  
We then summarize the paper in Sec. \ref{summary}. 

\section{The shift-invert Lanczos method for NEGF method}
\label{formulation}

\subsection{Non-equilibrium Green's function method}

Throughout this paper, we consider a Hamiltonian in a form of,
\begin{equation}
\label{Hamiltonian}
H  = \left(\begin{matrix}
     H_L & V_L & 0   \cr
     V^T_{L} & H_b & V_R\cr
     0 &  V^T_{R}& H_R   \cr
           \end{matrix}\right). 
\end{equation}
This type of Hamiltonian models systems in which 
the middle part of the Hamiltonian, $H_b$, is connected to the left and the 
right parts, $H_L$ and $H_R$, through $V_L$ and $V_R$, respectively.
In the case of electronic devices, $H_L$ and $H_R$ correspond to Hamiltonians for two 
electron reservoirs.
On the other hand, in nuclear fission, 
those correspond to Hamiltonians for a compound nucleus configuration and pre-scission configurations, 
respectively  \cite{Bertsch2023, Uzawa2024}.
We assume that there is no direct coupling between $H_L$ and $H_R$. 
Generally, the basis vectors are not orthogonal, and 
we also introduce the overlap matrix for the basis vectors,
\begin{equation}
    N_{i,j}=\langle i|j\rangle. 
\end{equation}
The resultant Schr\"odinger equation becomes a generalized 
eigenvalue equation, 
\begin{equation}
    Hf=ENf,
    \label{Eigen2}
\end{equation}
where $E$ and $f$ are eigenvalues and eigenvectors, respectively. 

When the system is coupled to external channels, one has to add self-energy terms  
to the Hamiltonian (\ref{Hamiltonian}) in order to describe decays of the system 
into the external channels. 
That is, $H\to H+\Delta-i\Gamma/2$, where 
$\Delta$ and $\Gamma$ are the real and imaginary parts of the self-energy originating 
from the couplings to the external channels.
The Hamiltonian matrix then becomes non-Hermitian, and currents are induced.
When there are two or more connections to external systems, one can calculate transition probabilities $T_{a,b}$
from a channel $a$ to a channel $b$ using the trace formula \cite{datta_1995, PRL1992, Miller1993},
\begin{equation}
\label{Datta}
    T_{a,b}(E)={\rm Tr}\left[\Gamma_{a}G(E)\Gamma_{b} G^{\dagger}(E)\right].
\end{equation}
Here $\Gamma_{a}$ and $\Gamma_{b}$ represent decay widths to the corresponding channels, 
and $G(E)$ is the retarded non-equilibrium Green function defined as,
\begin{equation}
\label{green}
    G(E)=\left[EN-\left(H+\Delta-\frac{i}{2}\Gamma\right)\right]^{-1},
\end{equation}
where $E$ is the energy of the system.  

The quantal trace formula (\ref{Datta}) has been 
widely applied to calculate the electronic properties of nano-devices \cite{datta_1995,Thoss2018} and 
cross sections of neutron induced fission \cite{Bertsch2023, Uzawa2024}. 
With the Green function $G(E)$, other quantities of a system, such as the spatial density distribution and the level density,  
can also be evaluated.

Notice that the Green function $G$ is the matrix inverse of the Hamiltonian, 
and therefore can be diagonalized by the same eigenstates of the Hamiltonian.
In the case of non-orthogonal basis, these eigenstates satisfy 
the equations 
\begin{equation}
    \left(H+\Delta-\frac{i}{2}\Gamma\right)f_{\lambda}=\tilde{E}_{\lambda}Nf_{\lambda},
        \label{right_eq}
\end{equation}
and 
\begin{equation}
    \tilde{f}^\dagger_{\lambda}\left(H+\Delta-\frac{i}{2}\Gamma\right)=\tilde{f}^\dagger_{\lambda}\tilde{E}_{\lambda}N.
        \label{left_eq}
\end{equation}
Here $\lambda$ is the label of eigenstates, $f_\lambda$ and $\tilde{f}_\lambda$ are the right-eigenvector and the 
left-eigenvector, respectively. 
$\tilde{E}_{\lambda}\equiv E_{\lambda}-\frac{i}{2}\Gamma_\lambda$ is a complex eigenvalue for $\lambda$.
Using these eigenstates, 
the matrix elements of the Green function are represented in a form of the spectrum decomposition as 
\begin{equation}
G_{ij}(E)
=\sum_\lambda\,\frac{(f_\lambda)_i\,(\tilde{f}_\lambda)_j^*}{E-\tilde{E}_\lambda}
=\sum_\lambda\,\frac{(f_\lambda)_i\,(\tilde{f}_\lambda)_j^*}{E-E_\lambda+i\Gamma_\lambda/2}.
\label{SD}
\end{equation}
This equation suggests that 
those eigenstates whose eigenvalue $E_{\lambda}$  
is close to the energy $E$ dominantly contribute to the Green function due to the energy denominator.
This motivates approximating 
the Green function by a selected set of the eigenstates. 

\subsection{Application of the shift-invert Lanczos algorithm}

In general, there is a large computational cost to invert a large Hamiltonian matrix.
In that circumstance, 
the spectral decomposition of the Green function 
may be useful by converting 
the matrix inversion problem into the 
eigenvalue problem. As we discussed in the previous subsection, one would need only 
a few eigenstates to represent the Green function. 
If the lowest or the highest eigenstates were necessary, one could have used the usual 
Lanczos algorithm. 
However, in the problem of NEGF, 
one needs eigenstates in the middle of the spectrum, and 
the direct application of the Lanczos algorithm is not efficient.
We thus introduce an alternative method with 
the shift-invert Lanczos algorithm. 

In the usual Lanczos algorithm, a symmetric or Hermitian matrix $H$ 
in an orthogonal basis (that is, $N$=1) 
is tridiagonalized within the following Krylov subspace \cite{Lanczos_text},
\begin{equation}
    {\rm span}\{\vec{q}, H\vec{q}, ..., H^{n-1}\vec{q}\}.
    \label{lanczos-vectors}
\end{equation}
Here $\vec{q}$ is an arbitrary initial vector and $n$ is the dimension of the subspace, which 
can be smaller than the dimension of the matrix $H$. 
The tridiagonalized matrix in the Krylov subspace reads,
\begin{equation}
\left(\begin{matrix}
     \alpha_0 & \beta_1 &  & &  \cr
     \beta_1 & \alpha_1 & \beta_2 &  & \cr
      &  \ddots& \ddots & \ddots&   \cr
      &  & \beta_{n-2}& \alpha_{n-2}& \beta_{n-1} \cr
      &  & & \beta_{n-1}& \alpha_{n-1} \cr
           \end{matrix}\right). 
\end{equation}
It is known that the eigensolutions of the tridiagonal matrix well approximate the eigensolutions of the full Hamiltonian,
and the eigensolutions are obtained 
in the order of the absolute value of eigenvalues 
\footnote{By adding an appropriate constant matrix to the Hamiltonian, the Lancozs algorithm always yields the ground state solution.}.

Naively, eigenstates in the middle of the spectrum may be obtained by  
first transforming the eigenvalue problem $H\psi=E\psi$ to 
\begin{equation}
   (H-\sigma)^2\psi=(E-\sigma)^2\psi,
\end{equation}
as suggested in Ref. \cite{Grosso1995}. 
Here $\sigma$ is a real parameter, and thereby the excited states of $H$ with the eigenvalue around $\sigma$ 
appear as the lowest eigenstates of $(H-\sigma)^2$. One can then apply the Lanczos algorithm to the 
matrix $(H-\sigma)^2$ 
to obtain 
those eigenstates. However, we found that convergence is sometime slow in this scheme.

We thus advocate using the shift-invert Lanczos algorithm \cite{Shift-Invert}. 
In this algorithm, 
the original generalized eigenvalue equation, Eq. (\ref{Eigen2}), is first 
transformed to
\begin{equation}
\label{SI1}
(H-\sigma N)f=(E-\sigma)Nf, 
\end{equation}
with a real parameter $\sigma$.
Then, applying  $(E-\sigma)^{-1}(H-\sigma N)^{-1}$ to the both sides of Eq. (\ref{SI1}), one obtains
\begin{equation}
    (H-\sigma N)^{-1}Nf=(E-\sigma)^{-1}f.
    \label{SI2}
\end{equation}
Here   
the original generalized eigenvalue problem is transformed into the eigenvalue problem of 
the matrix $(H-\sigma N)^{-1}N$, for which 
eigenvalues around $\sigma$ have large absolute values due to the form of $(E-\sigma)^{-1}$ on the right 
hand side of Eq. (\ref{SI2}). 
We solve this eigenvalue problem using the Lanczos iteration.
In the iteration, the matrix $(H-\sigma N)^{-1}N$ is applied to the Lanczos vectors, similar to 
Eq. (\ref{lanczos-vectors}) but with $(H-\sigma N)^{-1}N$ instead of $H$. 
The action of $N$ on the Lanczos vectors $\vec{v}$, $\vec{v}'\equiv N\vec{v}$, is the usual matrix multiplication.
In Eq. (\ref{SI2}), one needs to evaluate $\vec{x}\equiv (H-\sigma N)^{-1}\vec{v}'$. 
This is simplified by recognizing that 
one needs only the action of $(H-\sigma N)^{-1}$ onto the vector $\vec{v}'$ and the inverse of the 
matrix  $(H-\sigma N)$ itself does not need to be obtained. 
This can be done 
by firstly factorizing the matrix $(H-\sigma N)$ with e.g. LU factorization 
and then solving the linear equation,
\begin{equation}
    (H-\sigma N)\vec{x}= \vec{v}', 
\end{equation}
for $\vec{x}$. 
This method has been implemented in the package ARPACK \cite{ARPACK}, that is 
FORTRAN77 numerical software library for solving large-scale eigenvalue problems using the Arnoldi or Lanczos algorithms. 
In the next section, we demonstrate the effectiveness of this approach using a simple model Hamiltonian as well as 
a Hamiltonian for fission of $^{236}$U nucleus.

\section{results}
\label{results}

\subsection{GOE+barrier+GOE model\label{GOE}}

We first apply 
our method to 
a simple model Hamiltonian which has been 
used in the previous works to discuss the transition state dynamics with a potential barrier 
\cite{Weidenmuller2022, Hagino2023,Weidenmuller2024},
\begin{equation}
\label{Hamiltonian1}
H+\Delta  = \left(\begin{matrix}
     H_{1} & V_1 &0 \cr
     V_1^T & B_h & V_2^T \cr
       0& V_2 & H_{2} \cr
           \end{matrix}\right).
\end{equation}
Here, $H_{1}$ and $H_{2}$ are random matrices based on  
the Gaussian Orthogonal Ensemble (GOE) \cite{Brody1981,weidenmuller1,weidenmuller2}.
The GOE is characterized by two parameters, the dimension of the matrix $N_{\rm GOE}$ and the size of the matrix 
elements $v$. 
With these parameters, elements of a GOE matrix are given as \cite{Brody1981,weidenmuller1,weidenmuller2},  
\begin{equation}
\label{HGOE}
    (H_{\rm GOE})_{i,j}=vr(1+\delta_{i,j}), 
\end{equation}
where $r$ is a random number sampled from the standard normal distribution. 
In the calculations shown below, we take $N_{\rm GOE}=100$ 
and $v=0.1$ in the units of the typical energy scale of the system 
for both $H_1$ and $H_2$. In this subsection, all quantities related to energy are scaled in a similar way 
with the typical energy scale of the system.  

$B_h$ in Eq. (\ref{Hamiltonian1}) is a real scalar number, corresponding to the height of a barrier.
The GOE matrices and the barrier configuration are connected with vectors $V_1$ and $V_2$.
The elements of $V_1$ and $V_2$ are random numbers sampled from the normal distribution 
with the mean value of zero and the standard deviation of 0.1
\footnote{We have confirmed that our conclusion remains the same even when 
all the matrix elements of $V_1$ and $V_2$ are fixed to a constant number, $0.1$, 
although the convergence with respect to the number of eigenstates to be included 
becomes somewhat slower. 
}.
In this model, the basis vectors are orthogonal, and the overlap matrix $N$ becomes the unit matrix.

We assume that $H_1$ is connected to two different kinds of external channels, 
the input channel $\Gamma_{\rm in}$ and the left output channels $\Gamma_{{\rm L}}$. 
On the other hand, $H_2$ is assumed to connect to the right output channels $\Gamma_{{\rm R}}$. 
\footnote{In the problem of neutron induced fission reactions, $\Gamma_{\rm in}$, $\Gamma_{{\rm L}}$, and $\Gamma_{{\rm R}}$ correspond to a neutron channel, capture channels, and fission channels, respectively. } 
We assume that the decay matrices corresponding to these channels are given by, 
\begin{equation}
    (\Gamma_{\rm in})_{k,k'}=\gamma_{\rm in} \delta_{k,1}\delta_{k',1}, 
\end{equation}
\begin{equation}
\Gamma_{{\rm L}}  = \left(\begin{matrix}
     \gamma_{{\rm L}}I & 0 & 0 \cr
     0& 0 &  0 \cr
      0&  0& 0  \cr
      \end{matrix}\right), 
\end{equation}
and 
\begin{equation}
\Gamma_{{\rm R}}  = \left(\begin{matrix}
     0 & 0 & 0 \cr
     0& 0 &  0 \cr
      0&  0& \gamma_{{\rm R}}I  \cr
      \end{matrix}\right), 
\end{equation}
where $I$ is the unit matrix with the dimension of $N_{\rm GOE}$.
With this set up, the trace formula (\ref{Datta}) for the transition probabilities 
is transformed to  
\begin{equation}
T_{{\rm in,R}}=\gamma_{\rm in}\gamma_{{\rm R}}\sum_{j\in {\rm R}}|G_{1,j}|^2.
\label{Tnf}
\end{equation} 
and
\begin{equation}
T_{{\rm in,L}}=\gamma_{\rm in}\gamma_{{\rm L}}\sum_{j\in {\rm L}}|G_{1,j}|^2.
\label{Tnf2}
\end{equation} 
Using the spectrum decomposition of the Green function, 
the transition probability $T_{a,b}$ in these equations reads, 
\begin{align}
T_{a,b}
&=\gamma_{a}\gamma_{b}\sum_{\lambda} 
\frac{|(f_\lambda)_a|^2|(\tilde{f}_\lambda)_b|^2}{(E-E_\lambda)^2+(\Gamma_\lambda/2)^2} 
 \notag\\
&+\gamma_{a}\gamma_{b}\sum_{\lambda\neq\lambda'}
\frac{(f_\lambda)_a(f_{\lambda'})^*_a 
(\tilde{f}_\lambda)^*_b(\tilde{f}_{\lambda'})_b }{(E-\tilde{E}_\lambda)(E-\tilde{E}_{\lambda'})^*}.
\label{T_decomposition}
\end{align}

In the actual numerical calculations, we take 
$\gamma_{\rm in}=1.25\times10^{-3}$, $\gamma_{{\rm L}}=3.25\times10^{-3}$, and $\gamma_{{\rm R}}=1\times10^{-3}$, 
which are much smaller than the real parts of the Hamiltonian matrix elements, $v=0.1$ in Eq. (\ref{HGOE}). 
We take $B_h$=2 and set the excitation energy $E=0$.  
Since the values of $\gamma$ are small, one can perturbatively treat the decay matrix 
$\Gamma=\Gamma_{\rm in}+\Gamma_R+\Gamma_L$ 
to evaluate 
the eigenstaets of the non-Hermitian matrix $H+\Delta-i\Gamma/2$. 
That is, we first solve the Schr\"odinger equation without the imaginary part as
\begin{equation}
\label{pertubation1}
    (H+\Delta) f_\lambda=E_\lambda Nf_\lambda,
\end{equation}
and then evaluate the imaginary part of the eigenvalues approximately as 
\begin{equation}
\Gamma_{\lambda}=\sum_{i}
|(f_\lambda)_i|^2\Gamma_{ii}.
\label{gamma_perturbation}
\end{equation}
Notice that with this treatment  
the right-eigenvector $f$ and the left-eigenvector $\tilde{f}$ in Eqs. (\ref{right_eq}) and (\ref{left_eq}) 
do not need to be 
evaluated separately, because those coincide with each other for a Hermitian matrix, $H+\Delta$.  

\begin{figure}
\includegraphics[width=7cm]{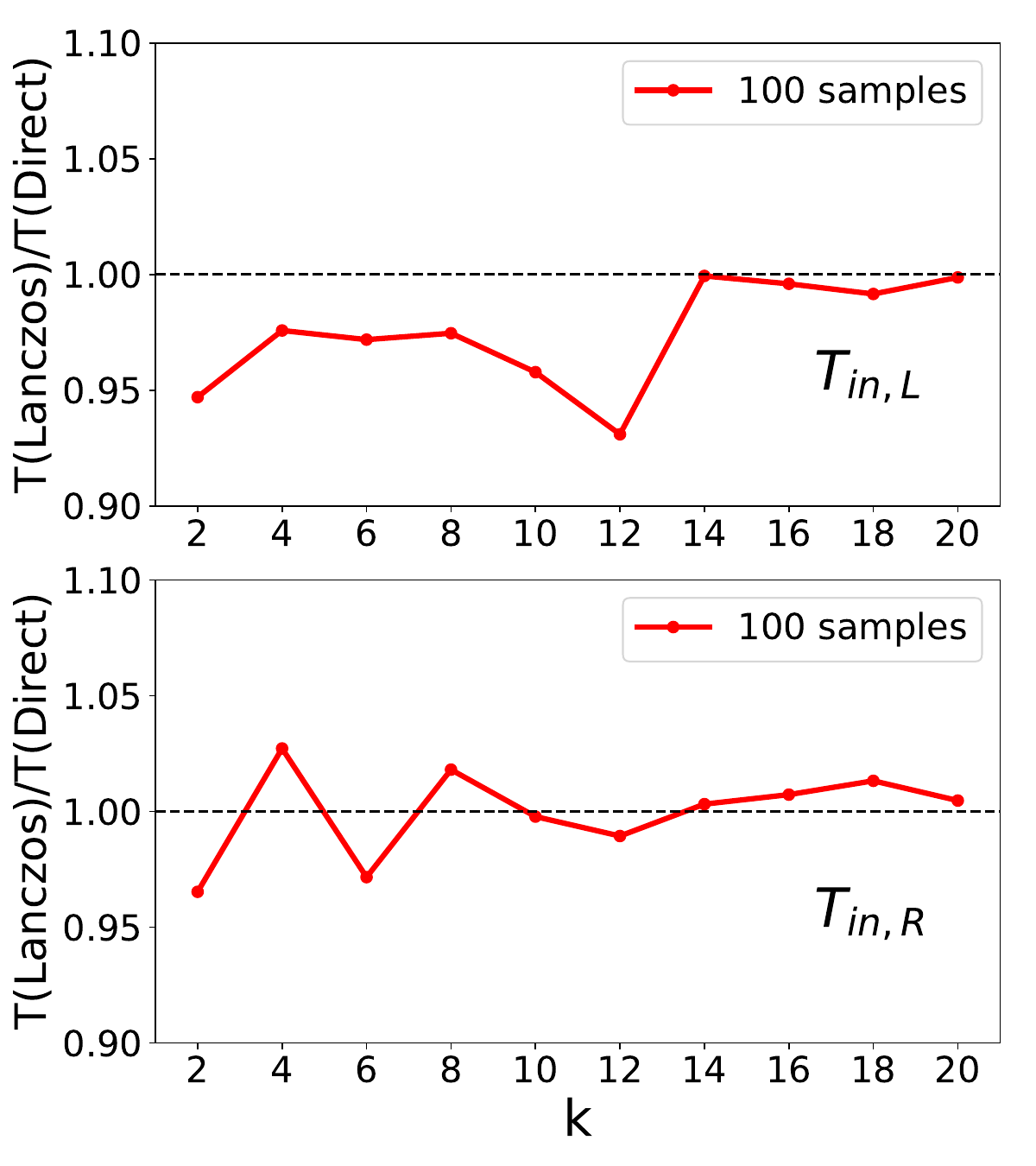}
\caption{
The ratio of transmission coefficients for the Hamiltonian
in Eq.(\ref{Hamiltonian1}) at $E=0$ with parameters defined in the text, calculated
by the two methods.  The number of eigenstates $k$ included in 
shift-invert Lanczos method is shown on the horizonal axis.
The results for $T_{{\rm in,L}}$  and $T_{{\rm in,R}}$ are shown in the upper and
lower panels, respectively.
}
\label{GOE1}
\end{figure}

The dimension of the Hamiltonian matrix in our calculation is 201, which is small enough 
to easily evaluate the Green function by the direct matrix inversion.
Let us then compare the transmission coefficients obtained by
the direct method to those with the 
shift-invert Lanczos method. 
For the latter calculations, 
we obtain $k$ eigenstates around the energy $E$ by setting $\sigma=E$, and 
use these $k$ eigenstates to calculate the transmission coefficients $T_{{\rm in,L}}$ and $T_{{\rm in,R}}$ 
by reducing the summation in Eq. (\ref{T_decomposition}) as  
\begin{equation}
\sum_{\lambda} \to \sum_{\lambda=1}^k. 
\end{equation}
Since the model uses the GOE ensemble, we need to average over sample matrices
to evaluate physical quantities. 
With our setup, we find that 100 samples provides sufficient convergence 
properties\footnote{We have confirmed that the results are not significantly altered by using other sets of 100 samples}. 
The ratios of the transition probabilities at $E=0$ calculated with 
the shift-invert Lanczos method to those with 
the direct method are shown in Fig.\ref{GOE1} as a function of the number 
of eigenstates $k$ obtained by the shift-invert Lanczos method. 
To this end, we follow  Ref. \cite{ARPACK} and randomly generate the elements of the initial vector $\vec{q}$ for the Lanczos algorithm. 
For each of the 100 samples, we use a different initial vector. 
The upper and the lower panels of Fig. \ref{GOE1} show the results for 
$T_{{\rm in,L}}$ and $T_{{\rm in,R}}$, respectively. 
In the cases with 1 sample (not shown), the transition coefficients 
obtained with the shift-invert Lanczos method do not converge even with large values of $k$.
As the sample number increases, the ratios become close to one. 
With 100 samples, the error becomes less than $1\%$ for both $T_{{\rm in,L}}$ and $T_{{\rm in,R}}$ at $k=20$.

\begin{figure}
\includegraphics[width=7cm]{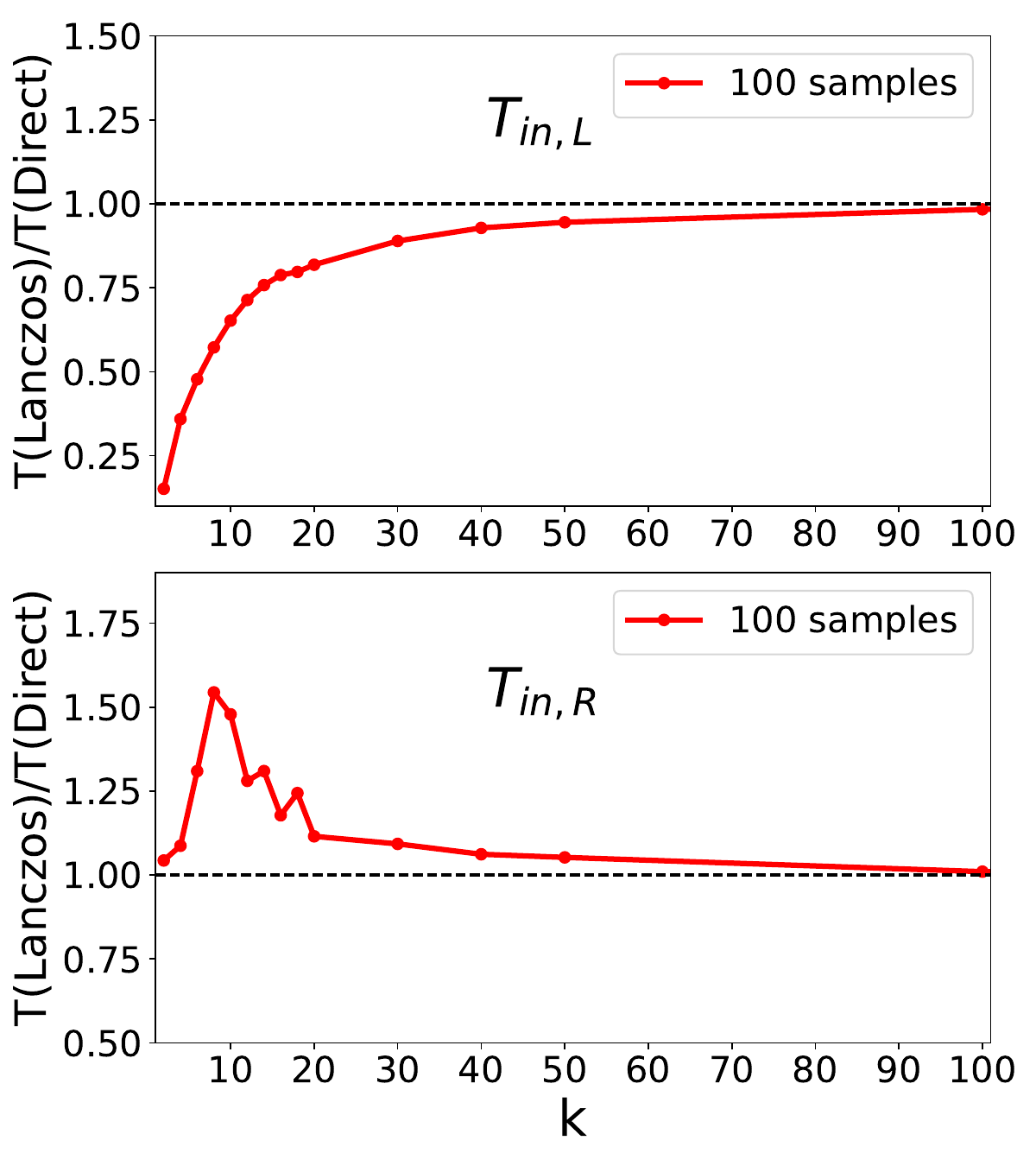}
\caption{
Same as Fig.\ref{GOE1}, but for larger values of the width parameters 
(see the main text).
The horizontal axis $k$ is the number of eigenstates obtained by the shift-invert Arnoldi method.} 
\label{GOE_Arnoldi}
\end{figure}

We next consider a case where 
$\gamma_{\rm in}$, $\gamma_{\rm L}$, 
and  $\gamma_{\rm R}$ are not small so that resonances are overlaping 
with each other. In this case, 
the approximation in Eqs. (\ref{pertubation1}) and (\ref{gamma_perturbation}) is not justified, and one needs to solve the eigenvalue equation for the Hamiltonian $H+\Delta-\frac{i}{2}\Gamma$ as it is. 
The Lanczos algorithm is no longer applicable to such non-Hermite Hamiltonian, 
and the Arnoldi algorithm, which is a generalization of the Lanczos algorithm, 
has to be applied \cite{ARPACK}. 
Figure \ref{GOE_Arnoldi} shows the results for $\gamma_{\rm in}=\gamma_{{\rm L}}=\gamma_{{\rm R}}=0.1$, 
while the other parameters are set to be the same as before. 
One can see that the transmission coefficients with the Arnoldi algorithm are converged to 
those with the direct calculations when the value of $k$ is large enough, even though 
a larger value of $k$ is required as compared to the case with isolated resonances shown in Fig. 1. 
The slower convergence can be easily understood with Eq. (\ref{T_decomposition}), 
which indicates 
that the more eigenstates contribute for the larger values of widths. 

\subsection{$^{236}$U with seniority zero configurations}

We next apply the same procedure to a more realistic case, that is, neutron induced fission 
of $^{236}$U nucleus. To this end, 
we calculate the Hamiltonian matrix of $^{236}$U based on the nuclear DFT \cite{DFT_text, Nakatsukasa2016,bender2003}. 
The employed Hamiltonian is the same as that used in the previous work\cite{Uzawa2024}, 
in which the details of the model Hamiltonian can be found. 

In the DFT calculation, we consider many-particle many-hole configurations at a given nuclear shape 
along a fission path. To generate such configurations, we first 
solve the Kohn-Sham equations by 
constraining the expectation value of the mass-quadrupole moment 
\begin{equation}
    Q=\int d\vecbold{r} \rho(\vecbold{r}) Y_{20}(\hat{\vecbold{r}}),
\end{equation}
where $\rho(\vecbold{r})$ is the mass-distribution of nucleons and $Y_{20}$ is the spherical-Harmonics function. 
In our calculations, the fission path is discretized such that the overlap between the neighboring 
configurations is $e^{-1}$\cite{Bertsch2023,Uzawa2024}. 
As a consequence, we have 14 reference points from the ground state deformation to a deformation prior to scission. 
The value of $Q$ for those reference points are summarized in Table 
\ref{dim1}. 
Notice that there are first and second barriers along the fission path, as shown in Fig. 1 
in Ref. \cite{Uzawa2024}.

At each reference point, 
we generate particle-hole excited configurations up to $E_{\rm max}=5$ MeV, for 
each of proton and neutron excitations. 
The dimension of a Hamiltonian at each $Q$ (i.e., \lq\lq $Q$-block") 
is also summarized in Table \ref{dim1}.
As in the previous subsection, 
we replace 
the left-most and the right-most blocks by GOE matrices $H^{(L)}_{\rm GOE}$ and $H^{(R)}_{\rm GOE}$. 
For the GOE parameters, we take $N_{\rm GOE}=1000$ and $v=0.32$ MeV.
The off-diagonal elements of the Hamiltonian matrix are evaluated 
according to a monopole pairing interaction and the diabatic interaction \cite{Uzawa2024}. 
We neglect the couplings among $Q$-blocks beyond the nearest neighboring,  and 
therefore the Hamiltonian matrix becomes block-tridiagonal as,
\begin{equation}
\label{3D-H0}
H  = \left(\begin{matrix}
     H^{(L)}_{\rm GOE} & V_L &  & & & \cr
     V^T_L & H_1 & V_{1,2} &  & \text{\large{\textit{O}}}& \cr
      & V_{2,1} & H_2 & V_{2,3} &  & \cr
      &  &  &\ddots &  & \cr
      \text{\large{\textit{O}}}&  & & V_{11,12}& H_{12} &V_R \cr
      &  & & & V^T_R &H^{(R)}_{\rm GOE} \cr
           \end{matrix}\right). 
\end{equation}
Here $H_i$ represents the Hamiltonian matrix at each quadrupole moment, $Q_i$, and 
$V_{i,i+1}$ represents the off-diagonal Hamiltonian matrix between the $Q$ blocks $i$ and $i+1$. 

\begin{table*}[htb] 
\caption{The quadrupole moment $Q$ and the dimension of each $Q$-block along a 
fission path of $^{236}$U. The quadrupole moment is given in units of barn. 
The total dimension is 66103 with $N_{\rm GOE}=1000$. 
\label{dim1}}
\begin{tabular}{c|cccccccccccccc}
\hline 
\hline 
$Q$(barn)  & 14&18&  23 & 29 & 34 & 39 & 46 & 51 & 57 &62 & 67 & 74 & 79 &83\\
\hline 
 dimension& $N_{\rm GOE}$&2520&  9794 & 15088 & 11577 & 2774 & 2940 & 3021 & 3150 & 2196 & 3752 & 2871 & 4420 &$N_{\rm GOE}$\\
\hline 
\hline 
\end{tabular}

\end{table*}

Similarly to the previous subsection, based on the shift-invert Lanczos method 
for the trace formula, Eq. (\ref{Datta}), 
we calculate the transition probabilities 
$T_{n,\rm cap}$ and $T_{n,\rm fis}$ at $E=$ 6.5 MeV, 
which is equal to the neutron separation energy in $^{236}$U \cite{Leal1999}. We use the same 
partial widths to those in Sec. \ref{GOE}, 
for which 
$\Gamma_{\rm in}$, $\Gamma_{\rm L}$ and $\Gamma_{\rm R}$ correspond to 
$\Gamma_n$, $\Gamma_{\rm cap}$ and $\Gamma_{\rm fis}$, respectively.
Following Ref. \cite{Uzawa2024}, we set $\gamma_{n}=0.01$ MeV, $\gamma_{\rm cap}=0.00125$ MeV, and $\gamma_{\rm fis}=0.015$ MeV.
Similarly to the previous subsection, $\Gamma_{n}$ has one diagonal component in the left-most block, 
while $\Gamma_{\rm cap}$ and $\Gamma_{\rm fis}$ have $N_{\rm GOE}$ 
diagonal components in the left-most and the right-most blocks, respectively. 
In the shift-invert Lanczos calculation, we set $\sigma=6.5$ MeV.  

\begin{figure}
\includegraphics[width=7cm]{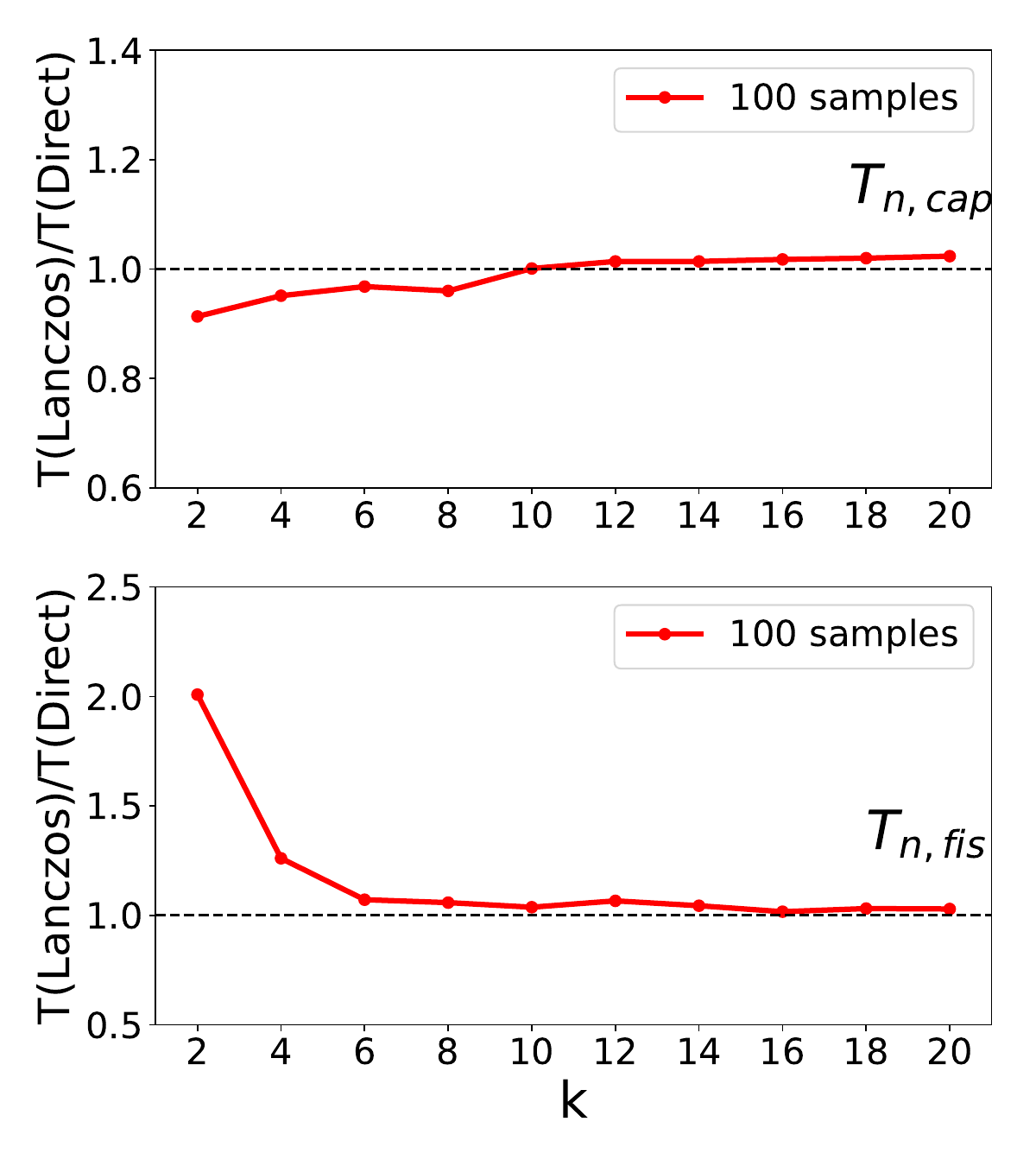}
\caption{
Similar to Fig.\ref{GOE1}, but for the transition probabilities at $E$=6.5 MeV 
for fission of $^{236}$U.  
For the shift-invert Lanczos method, we set $\sigma=6.5$ MeV.
}
\label{236U1}
\end{figure}

Fig. \ref{236U1} shows the ratios of the transition probabilities calculated with the shift-invert Lanczos method to 
those with the direct method. 
The calculations were performed using a single node equipped in Yukawa-21 supercomputer at the Yukawa Institute for 
Theoretical Physics (YITP), Kyoto University. 
We apply the ZGETRF and ZGETRI subroutines included in Lapack \cite{laug} for the calculations 
of the matrix inversion. On the other hand, the ARPACK library is used for the 
calculation of shift-invert Lanczos Algorithm \cite{ARPACK}. 
By taking an ensemble average with 100 samples, the error of our approach is smaller 
than 3\% for both capture and fission, as shown by the red lines in the figure. 
Moreover, one can also observe that 
the results are almost converged with $k=10$. 
That is, we need only as small as 10 eigenstates to represent the Green function, whose dimension is 66103. 

\begin{figure}
\includegraphics[width=7cm]{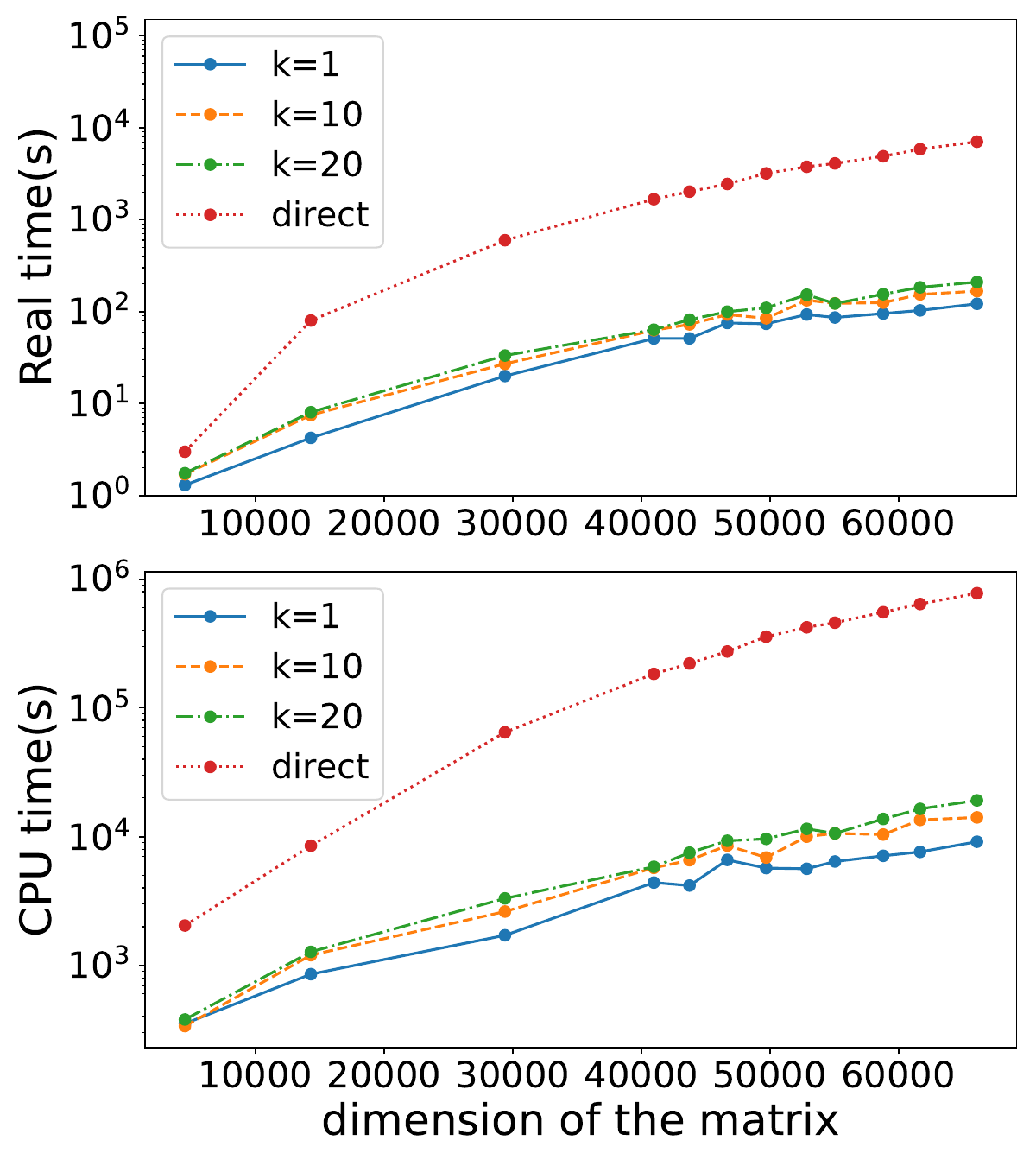}
\caption{
(The upper panel)
The real computational time for the shift-invert Lanczos method and the direct matrix inversion.
To draw this figure, 
the left-end and the right-end GOE matrices are fixed and the number of intermediate Q-blocks is increased one by one.
The quantity $k$ in the legend denotes the number of eigenvectors calculated with the shift-invert 
Lanczos method.
See the text for details of the numerical calculations.
(The lower panel) Similar to the upper panel, but for the CPU times.
}
\label{time}
\end{figure}

\begin{table*}[htb] 
\caption{
The computational times required to compute the Green function with the direct matrix inversion method 
and with 
the shift-invert Lanczos method.
\label{table:time}}
\begin{tabular}{c|c|c|c|c|c}
  \hline
  \hline
  &\multirow{2}{*}{dimension} &\multirow{2}{*}{direct}  & \multicolumn{3}{c}{shift-invert Lanczos} \\ 
  \cline{4-6}
  &   & &$k$=1& $k$=10& $k$=20 \\ 
  \hline
  \multirow{3}{*}{Real time(s)}&4520 & 3.59 & 1.18& 1.51& 1.89 \\
  &43753 & $2.01\times10^3 $ & $5.11\times10$& $7.26\times10$& $8.17\times10$ \\
  &66103 (full) & $7.03\times10^3 $ & $1.22\times10^2 $& $1.67\times10^2$& $2.10\times10^2$ \\
  \hline
  \multirow{3}{*}{CPU time(s)}& 4520 & $2.04\times10^3$ & $3.35\times10^2 $ & $3.42\times10^2 $ & $3.81\times10^2 $ \\
  &43753 & $2.21\times10^5 $ & $4.17\times10^3$ & $6.57\times10^3$ & $7.51\times10^3$ \\
  &66103 (full) & $7.77\times10^5 $ & $9.14\times10^3$& $1.41\times10^4$& $1.93\times10^4$ \\
  \hline
  \hline
\end{tabular}

\end{table*} 
 
Let us next discuss computation times needed 
to compute the Green function for fission of $^{236}$U, 
\begin{equation}
    G(E)=\left(EN-(H-\frac{i}{2}\Gamma)\right)^{-1}. 
    \label{direct}
\end{equation}
In the following calculations, we fix the left-end and the right-end GOE matrices 
and increase the number of the intermediate subblocks $H_i$ and  $V_{i,i+1}$ one by one.
The real calculation times and CPU times needed for the calculations 
are shown in the upper and the lower panels in Fig. \ref{time} 
as a function of the matrix dimension, respectively. 
As shown in the upper panel, the shift-invert Lanczos method 
reduces the real computational time in all the cases. 
For the dimension of 66103, 
the shift-invert Lanczos method is faster by a factor of about 33.4 as compared to the 
direct matrix inversion.  
The result is similar for the CPU time shown in the lower panel, although the ratio of the CPU times is 
increased to about 40.3. 
The computation times 
are summarized in Table \ref{table:time}.  
Notice that 
the computational cost of the direct matrix inversion 
increases as $O(N^3)$, while 
that for the shift-invert Lanczos method changes much weakly. 
Therefore, the shift-invert Lanczos method provides a powerful tool 
for transport phenomena, especially when the dimension of a Hamiltonian is large.

\section{summary}
\label{summary}

We have discussed the applicability of the shift-invert Lanczos approach for the NEGF method.
In the shift-invert Lanczos approach, 
the Green function is expressed in 
a form of the spectrum decomposition with 
eigenvalues and eigenstates of a generalized eigenvalue equation.
As only those eigenstates around $E$ contribute to transition probabilities, 
the shift-invert Lanczos approach can be applied to obtain such eigenstates efficiently. 
We have demonstrated that 
such approach 
reproduces the exact result by the direct inversion 
of a matrix with good accuracy for both a simple schematic model and for a more realistic case of fission of $^{236}$U nucleus. 
We have also demonstrated that 
the computation time can be considerably reduced, e.g., by a factor of about 30 for the fission problem of $^{236}$U. 

For the problem of fission, so far we have restricted the model space only to  
seniority zero configurations, that is, all nucleons are assumed to form time-reversal pairs. 
In more realistic calculations, it is essential to include also 
finite seniority configurations. 
A serious problem towards such calculations is that 
the dimension of a Hamiltonian matrix can be as huge as $O(10^6)$ \cite{Bertsch2023}.
In such occasion, 
the shift-invert Lanczos approach discussed in this paper shall provide a promising tool.

\section*{Acknowledgments}

We thank G.F. Bertch for useful discussions and for a careful reading of the manuscript. 
This work was supported in part by JSPS KAKENHI
Grant No. JP23K03414 and JP23KJ1212. 
The numerical calculations were performed
through the use of SQUID at the Cybermedia Center, Osaka University and Yukawa-21 at Yukawa Institute for Theoretical Physics, Kyoto University.

\bibliography{Lanczos}

\end{document}